\documentclass[aps,preprint]{revtex4}
\usepackage{amsfonts}
\usepackage{subfigure}
\usepackage{amsmath}
\usepackage{amssymb}
\usepackage{graphicx}

\input{tcilatex}

\begin{document}

\preprint{}
\title[ ]{Environment generated quantum correlations in bipartite
qubit-qutrit systems}
\author{Salman Khan}
\email{sksafi@comsats.edu.pk}
\author{Ishaq Ahmad}
\affiliation{Department of Physics, COMSATS Institute of Information Technology, Chak
Shahzad, Islamabad, Pakistan.}
\date{May 18, 2015}

\begin{abstract}
The dynamics of entanglement and quantum discord for qubit-qutrit systems
are studied in the presence of phase damping and amplitude damping noises.
Both one way and two couplings of the marginal systems with the environments
are considered. Entanglement sudden death is unavoidable under any setup,
however, the required time span depends on the way of coupling. On the other
hand, the dynamics of quantum discord strongly depends both on the nature of
environment and on the number of dimensions of the Hilbert space of the
coupled marginal system. We show that freezing and invariance of quantum
discord, as previously reported in the literature, are limited to some
special cases. Most importantly, it is noted that under some particular
coupling the existence of environment can guarantee the generation of
nonclassical correlations.


\end{abstract}

\maketitle

\section{Introduction}

Quantum mechanics is one of the theories that is more perplexing and yet
very fascinating which provides a complete physical description to
fundamental phenomenon taking place at the atomic levels. Indeterminism and
nonlocality are the two fundamental and controversial concepts of quantum
theory. In quantum theory the sources of nonlocality are Aharonov-Bohm
effect and entanglement. Entanglement has been recognized as the first
candidate of non-classical correlations and is considered as a vital
resource for quantum information science \cite{Horodecki1}. It has been
widely investigated and a number of schemes have been presented for its
detection, quantification and applications \cite{Bennett,Horodecki2,
Wootters}.

A generic quantum state $\rho$ is an object which is characterized not only
by quantum but also by classical correlations. From quantum information
point of view, distinguishing these two types of correlations is of utmost
importance. One way to do this is to use entanglement versus separability
formalism introduced by Werner \cite{Werner}. Moreover, in a quantum state
there exist correlations that are not captured by the measures of
entanglement and are still useful in carrying out certain quantum
information tasks. Quantum discord, with no classical analog, introduced by
Ollivier and Zurek \cite{Zurek} and, independently, by Henderson and Vedral 
\cite{Henderson} for bipartite systems is a more general measure of quantum
correlations which also captures those that do not come in the domain of
quantum entanglement. It has been used as a resource for certain quantum
computation models \cite{Datta}, encoding of information onto a quantum
state \cite{Gu} and quantum state merging \cite{Madhok, Cavalcanti}. Being
an important measure of quantum correlations, the domain of the measurement
of quantum discord has been recently extended to continuous variable systems
of Gaussian and non-Gaussian states \cite{Adesso,Giorda,Tatham}. The study
of the behavior of quantum discord for a two mode squeezed state in
noninertial frames shows that quantum discord asymptotes to zero in the
limit of infinite acceleration \cite{Doukas}. A detail of other studies
related to quantum discord, alternative measures of quantum correlations and
its behavior in different setups are discussed in Refs. [18-26].

The irreversible loss of fundamental quantum features such as quantum
superposition by a quantum system when it interacts with an environment is
one of the big issues in the practical implementation of different protocols
based on these quantum features. The utility of entanglement in different
quantum information tasks as a resource decreases when the system interacts
for considerably long time with its environment. The effect of environment
on entanglement between the components of a composite system leads to a
number of undesirable consequences, such as entanglement loss, entanglement
sudden death and even the rebirth of entanglement \cite{Eberly, Salman,
sharma, Nayak}. The study of the dynamics of quantum entanglement and
quantum discord in Markovian environment shows that, unlike entanglement,
quantum discord is immune to sudden death \cite{Werlang1}. A sudden
transition between classical and quantum loss of correlations for a class of
Bell diagonal states under local dephasing noise has also been reported \cite%
{Mazzola}. This behavior predicts that there exists a finite time interval,
in which only classical correlations decay and quantum discord is frozen
despite the presence of a noisy environment. In Ref. \cite{Haikka} the
authors have shown that depending on the initial state, the quantum discord,
in the presence of non-Markovian purely dephasing environment, might get
frozen forever at a positive value. The behavior of quantum discord and
entanglement in the presence of local qutrit dephasing noise is studied in
Ref. \cite{Karpat1,Karpat2}. It is shown that the entanglement for the
system vanishes in a finite time interval but the quantum discord remain
invariant throughout the time evolution.

In this paper, we study the behaviors of entanglement and quantum discord
for a particular bipartite state, which consists of a qubit and a qutrit as
marginal systems where each interacts with a local environment. The
influences of two types of noisy environments, the dephasing and the
amplitude damping noises are investigated. Both one way and two ways
coupling of the marginal systems in the form of local and multilocal
environments are considered. We find that the freezing and decay, as
reported previously, in the presence of a noisy environment are not the only
aspects of quantum discord. It is found that the behavior of quantum discord
strongly depends on the dimensions of the Hilbert space of the marginal
system. We show that maneuvering the type of coupling between the system and
a particular environment can lead to the generation or destruction of
quantum discord.

\section{Measures of quantum correlations}

In this section we briefly review the quantifiers for entanglement and non
classical correlations. Many entanglement measures for quantifying
entanglement of bipartite states exist in the literature. However, we will
use negativity which is a reliable measure of entanglement of bipartite
states of any dimensions, provided that the state has a negative partial
transpose. The partial transpose of a bipartite density matrix $%
\rho_{m\nu,n\mu}$ over the second qubit $B$ is given by $\rho_{m\mu,n%
\nu}^{T_{B}}=\rho_{m\nu,n\mu}$ and for the first qubit, it can similarly be
defined. For a bipartite state $\rho^{AB}$, the negativity $\mathcal{N}%
(\rho^{AB})$ is defined as twice the absolute sum of the negative
eigenvalues of partial transpose of $\rho^{AB}$ with respect to the smaller
dimensional system,%
\begin{equation}
\mathcal{N}(\rho^{AB})=\sum_{i}\left\vert \lambda_{i}\right\vert
-\lambda_{i},  \label{E1}
\end{equation}
where $\lambda_{i}$ are the eigenvalues of the partial transposed density
matrix.

The nonclassical correlation are quantified by discord. The discord $%
\mathcal{D}(\rho^{AB})$ for a bipartite state $\rho^{AB}$ is defined as the
difference between total correlations $I(\rho^{AB})$ and the classical
correlation $C(\rho^{AB})$,%
\begin{equation}
\mathcal{D}(\rho^{AB})=I(\rho^{AB})-C(\rho^{AB}).  \label{E2}
\end{equation}
The quantum mutual information $I(\rho^{AB})$ is a measure of total amount
of classical and quantum correlations in a quantum state. Mathematically, it
is given by%
\begin{equation}
I(\rho^{AB})=S\left( \rho^{A}\right) +S\left( \rho^{B}\right) -S\left(
\rho^{AB}\right) ,  \label{E3}
\end{equation}
where $S(\rho)=-Tr\left( \rho\log_{2}\rho\right) $ is the von Neumann
entropy of the system in the state $\rho$ and $\rho^{A(B)}=Tr_{B(A)}\left(
\rho^{AB}\right) $ are the two marginal states of the composite system. The
classical correlation by definition is the maximal information that one can
obtain and is mathematically given by \cite{Zurek,Henderson}%
\begin{equation}
C_{B}(\rho^{AB})=S\left( \rho^{B}\right) -\min_{\{\Pi_{k}^{A}\}}\sum
_{k}p_{k}S\left( \rho_{k}^{B}\right) ,  \label{E4}
\end{equation}
where $\rho_{k}^{B}=Tr_{A}\left( \left( \Pi_{k}^{A}\otimes I^{B}\right)
\rho^{AB}\left( \Pi_{k}^{A}\otimes I^{B}\right) \right) /p_{k}$ is the
postmeasurement state of subsystem $B$ after obtaining the outcome $k$ on
subsystem $A$ with probability $p_{k}=Tr\left( \left( \Pi_{k}^{A}\otimes
I^{B}\right) \rho^{AB}\left( \Pi_{k}^{A}\otimes I^{B}\right) \right) $. The
set $\left\{ \Pi_{k}^{A}\right\} $ are projectors onto the space of marginal
state $A$ and $I^{B}$ is the identity operator for the space of marginal
state $B$ of the composite system. It is important to mention that for a
general mixed state, the measure of classical correlations is not symmetric,
that is, $C_{A}(\rho^{AB})\neq C_{B}(\rho^{AB}),$ as a result, the quantum
discord depends on which marginal state the projective measurement is
carried on. However, it is known that $\mathcal{D}_{A}(\rho^{AB}),\mathcal{D}%
_{B}(\rho^{AB})\geq0$ and $\mathcal{D}_{A}(\rho^{AB})=\mathcal{D}%
_{B}(\rho^{AB})=0$ if and only if $\rho^{AB}$ is a classical-quantum state.
We will investigate the quantity $C_{B}(\rho^{AB})$ for qubit-qutrit states
such that the projective measurement is made on the qubit marginal state of
the composite system. The measurement operators $\Pi_{k}$ ($k=1,2$) in the
qubit space can be expressed as follows%
\begin{equation}
\Pi_{k}=\frac{1}{2}\left( I\pm\sum_{j}n_{j}\sigma_{j}\right) ,  \label{E5}
\end{equation}
where the $\pm$ sign corresponds to $k=1,2$, respectively and $\sigma_{j}$ ($%
j=1,2,3$) are the three Pauli spin matrices. The vector $n$ defines a unit
vector on Bloch sphere having components $n=(\sin\theta\cos\phi,\sin\theta
\sin\phi,\cos\theta)^{T}$ with $\theta$ $\in$ $\left[ 0,\pi\right] $ and $%
\phi$ $\in$ $\left[ 0,2\pi\right] $. The main obstacle in obtaining quantum
discord for a general quantum state lies in the minimization procedure,
which is taken over all possible von Neumann measurements, of the quantum
conditional entropy of equation (\ref{E4}). The analytical expressions for
classical correlation and quantum discord are only available for two-qubit
Bell diagonal state and a seven-parameter family of two-qubit X states \cite%
{Luo, Ali} till now. For the simple qubit--qutrit states that we consider in
our work, we will calculate the quantum discord via numerical minimization
over the two independent real parameters $\theta$ and $\phi$.

\section{The system in noisy environment}

We consider a composite system of a qubit $A$ and a qutrit $B$ such that the
two marginal systems are locally or multilocally coupled to their
environments. The local and multilocal couplings describe the situations
when either the qubit or the qutrit or both the qubit and the qutrit are
independently influenced by their own environments.

The interaction of a system with an environment is studied in terms of
various quantum channels (noise) such as phase damping noise and amplitude
damping noise. When the density matrix of a system is influenced by a
dephasing noise, the diagonal elements of the density matrix remain
unaffected while the off-diagonal elements decay. The dynamics of quantum
correlations of the system we have chosen for the present studies have been
investigated in Ref. \cite{Karpat} under the condition that only the qutrit
is coupled locally to dephasing environment. In the first part of our study,
we reconsider the effect of dephasing noise on the same system, however, in
multilocal coupling and show that such coupling of the system with dephasing
noise can result in generation of quantum discord and also can leave it
noninvariant. The second part of our study explores the influence of
amplitude damping noise on the dynamics of quantum correlations. It is shown
that for a finite time both local and multilocal coupling with amplitude
damping noise generate quantum discord, however, completely destroys it in
the asymptotic limit. The easy way of studying the dynamics of an open
quantum system is to use Kraus operator formalism. The Kraus operators for a
single qubit and single qutrit dephasing noise are, respectively, given as%
\begin{equation}
E_{A0}^{D}=\text{diag}\left( 1,\sqrt{1-\gamma_{A}(t)}\right) ,\qquad
E_{A1}^{D}=\text{diag}\left( 0,\sqrt{\gamma_{A}(t)}\right) ,  \label{E6}
\end{equation}

\begin{align}
E_{B0}^{D}& =\text{diag}\left( 1,\gamma _{B}(t),\gamma _{B}(t)\right)
,\qquad E_{B1}^{D}=\text{diag}\left( 0,\sqrt{1-\gamma _{B}^{2}(t)},0\right) ,
\notag \\
E_{B2}^{D}& =\text{diag}\left( 0,0,\sqrt{1-\gamma _{B}^{2}(t)}\right) ,
\label{E7}
\end{align}%
where the time dependent parameters are given by $\gamma
_{A}(t)=1-e^{-t\Gamma _{A}}$ and $\gamma _{B}(t)=1-e^{-t\Gamma _{B}}$. The $%
\Gamma _{i}$ ($i=A,B$) represent the decay rates of the two marginal
systems. The Kruas operators for both single qubit and single qutrit
amplitude damping noise are not all diagonal and are given by%
\begin{equation}
E_{A0}^{A}=\left( 
\begin{array}{cc}
1 & 0 \\ 
0 & \sqrt{1-\beta _{A}(t)}%
\end{array}%
\right) ,\qquad E_{A1}^{A}=\left( 
\begin{array}{cc}
0 & \sqrt{\beta _{A}(t)} \\ 
0 & 0%
\end{array}%
\right) ,  \label{E8}
\end{equation}%
\begin{align}
E_{B0}^{A}& =\left( 
\begin{array}{ccc}
1 & 0 & 0 \\ 
0 & \sqrt{1-\beta _{B}(t)} & 0 \\ 
0 & 0 & \sqrt{1-\beta _{B}(t)}%
\end{array}%
\right) ,\qquad E_{B1}^{A}=\left( 
\begin{array}{ccc}
0 & \sqrt{\beta _{B}(t)} & 0 \\ 
0 & 0 & 0 \\ 
0 & 0 & 0%
\end{array}%
\right) ,  \notag \\
E_{B2}^{A}& =\left( 
\begin{array}{ccc}
0 & 0 & \sqrt{\beta _{B}(t)} \\ 
0 & 0 & 0 \\ 
0 & 0 & 0%
\end{array}%
\right) ,  \label{E9}
\end{align}%
where the time dependent parameters $\beta _{i}(t)$ are defined similar to
the $\gamma _{i}$'s along with decaying parameters as above. It is easy to
see that both $\gamma _{i},\beta _{i}$ $\epsilon $ $\left[ 0,1\right] $,
where the lower limit corresponds to no coupling and the upper limit
corresponds to complete coupling of the system and environment. The Kraus
operators for both qubit and qutrit satisfy the completeness relation $%
\sum_{i}E_{i}^{\dag }E_{i}=I$. The evolution of the initial density matrix
of the system when it is influenced by the multilocal noise is given in the
Kraus operators formalism as follows%
\begin{equation}
\rho (t)=\sum_{j=0}^{3}\sum_{k=0}^{2}\left( E_{Bj}E_{Ak}\right) \rho
(0)\left( E_{Ak}^{\dag }E_{Bj}^{\dag }\right) ,  \label{E10}
\end{equation}%
where $E_{Ak}=E_{Am}\otimes I_{3}$, $E_{Bj}=I_{2}\otimes E_{Bn}$ are the
Kraus operators of the local coupling of the qubit and the qutrit
individually. The system is said to be coupled multilocally when the
marginal states are influenced simultaneously by their local environments.
The subscripts $m=0,1$, and $n=0,1,2$ stand, respectively, for a single
qubit and a single qutrit Kraus operators of dephasing or amplitude damping
noise. The identity matrices $I_{2}$ and $I_{3}$ act, respectively, on the
qubit and qutrit parts of the composite system. we consider the initial
density matrix $\rho (0)$ of the composite system given by the following one
parameter family of matrices%
\begin{align}
\rho (0)& =\frac{p}{2}(|00\rangle \langle 00|+|01\rangle \langle
01|+|12\rangle \langle 12|+|11\rangle \langle 11|+|01\rangle \langle 11| 
\notag \\
& +|11\rangle \langle 01|+|00\rangle \langle 12|+|12\rangle \langle 00|)+%
\frac{1-2p}{2}(|02\rangle \langle 02|  \notag \\
& +|02\rangle \langle 10|+|10\rangle \langle 02|+|10\rangle \langle 10|),
\label{E11}
\end{align}%
where $p\in \left[ 0,0.5\right] $ and $\rho (0)$ is separable only when $%
p=1/3$.

We first analyze the case when only the qubit is locally coupled to the
phase damping noise. Under this condition, the nonzero matrix elements of $%
\rho(t)$ are given by%
\begin{align}
\rho_{11} & =\rho_{22}=\rho_{55}=\rho_{66}=\frac{p}{2},  \notag \\
\rho_{33} & =\rho_{44}=\frac{1-2p}{2},  \notag \\
\rho_{16} & =\rho_{25}=\rho_{52}=\rho_{61}=\frac{pe^{-t\Gamma_{A}/4}}{2}, 
\notag \\
\rho_{34} & =\rho_{43}=\frac{(1-2p)e^{-t\Gamma_{A}/4}}{2}.  \label{E12}
\end{align}
It can be seen from the above matrix elements that only the offdiagonal
elements decay with time. To observe the behavior of entanglement, we take
partial transpose of $\rho(t)$ over the qubit and using Eq. (\ref{E1}), the
negativity becomes%
\begin{equation}
\mathcal{N}(\rho)=\frac{1}{2}e^{-\Gamma_{A}/4}((1-e^{\Gamma_{A}/4})(1-p)+%
\left\vert (2+e^{\Gamma_{A}/4})p-1\right\vert +\left\vert
p-e^{\Gamma_{A}/4}(1-2p)\right\vert ).  \label{E13}
\end{equation}
The quantum mutual information is given by%
\begin{align}
I(\rho) & =\frac{1}{\log4}[2e^{-t\Gamma_{A}/4}(\tanh^{-1}(e^{-t%
\Gamma_{A}/4})-p\tanh^{-1}(pe^{-t\Gamma_{A}/4}))  \notag \\
& +4\tanh^{-1}(\frac{p}{3p-2})-p(\log4+4\log(1-2p)  \notag \\
& -2\log(p+p^{2}))+\log(4+\frac{4(p^{2}-1)}{e^{-t\Gamma_{A}/2}-p^{2}})].
\label{E14}
\end{align}
The analytical relation for quantum conditional entropy is very large, which
results in very lengthy and complicated relation for classical correlation $%
C_{B}(\rho)$. This in turn gives a complicated relation for quantum discord
which is impossible to be optimized analytically. Therefore, we will resort
to numerical optimization for studying the behavior of quantum discord for
all cases of coupling between the system and environment that we consider in
this paper.

\section{Numerical Analysis}

\begin{figure}[h]
\begin{center}
\subfigure[]{
\includegraphics[scale=1.05]{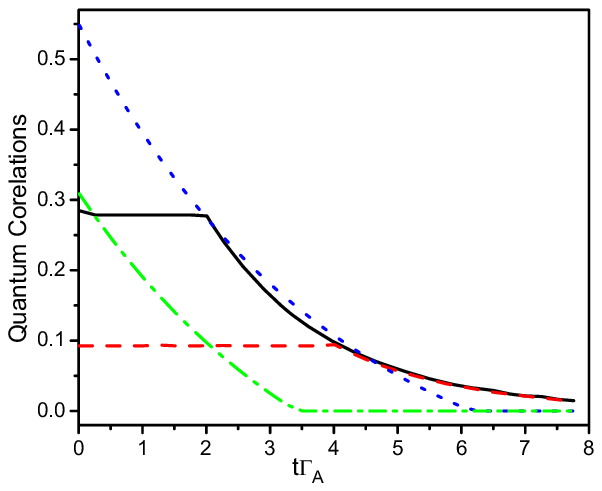}} 
\subfigure[]{
\includegraphics[scale=1.05]{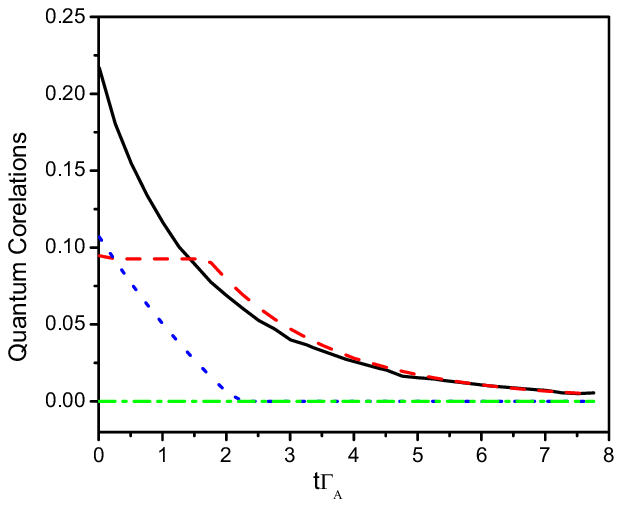}}
\end{center}
\caption{(Color Online) Both the negativity (blue dotted, green dash-dotted)
and the quantum discord (black solid, red dashed) are plotted, respectively,
for two values of $p=(0.15,\hspace{2mm}0.23)$ against the dimensionless
parameter $t\Gamma _{A}$. (a) Shows the dynamics of correlations when only
the qubit is locally coupled to dephasing noise. (b) Shows the situation for 
$t\Gamma _{B}=2$ when both the marginal systems are coupled locally to
dephasing noise.}
\label{Figure1}
\end{figure}

For the purpose of numerical optimization, we first study the case when only
the qubit is locally coupled to dephasing environment. Figure \ref{Figure1}$%
(a)$ shows the behaviors of negativity and quantum discord against the
dimensionless parameter $t\Gamma _{A}$ for two values of the state parameter 
$p=0.15,0.23$. It can be seen that for both choices of $p$, the decoherence
disentangles the system in finite time, however, the time domain in which
the complete loss of entanglement occurs is different for each value of $p$.
Comparatively, it survives longer (blue (dotted) curve) for $p=0.15$ than
for $p=0.23$ (green (dash-dotted) curve). Although quantitatively different,
this behavior of entanglement is qualitatively similar to what is reported
in the case of local coupling of the qutrit with dephasing environment \cite%
{Karpat2}. On the other hand, the dynamics of quantum discord strongly
depend on which marginal system is locally coupled to dephasing environment.
Unlike the results of Ref. \cite{Karpat2}, which shows invariance quantum
discord for $p=0.23$, we find that for both values of $p$ the decay in
quantum discord (black (solid) and red (dashed) curves) happens, however, at
different instants of time. The time span for which discord remains freezed
increases with the value of $p$. The remarkable aspect is that there exists
a critical value of time at which quantum discord becomes independent of the
state parameter $p$. The dynamics of quantum correlations of the system
under multilocal coupling with dephasing environment are shown in figure \ref%
{Figure1}$(b)$. The blue (dotted $p=0.15$) and the green (dashed-dotted $%
p=0.23$) curves represent the behavior of entanglement. We see that for
small values of $p$ the entanglement sudden death occurs very quickly and
there is no entanglement at all for large values of $p$. Regardless of the
value of $p$, the discord (black (solid) and red (dashed) curves) is always
larger than the entanglement and exists in the system beyond the instant at
which entanglement loss has taken place. The solid curve ($p=0.15$) shows
that in the presence of multilocal environment the discord monotonically
decreases without being freezed for any interval.

\begin{figure}[h]
\begin{center}
\includegraphics[scale=1.5]{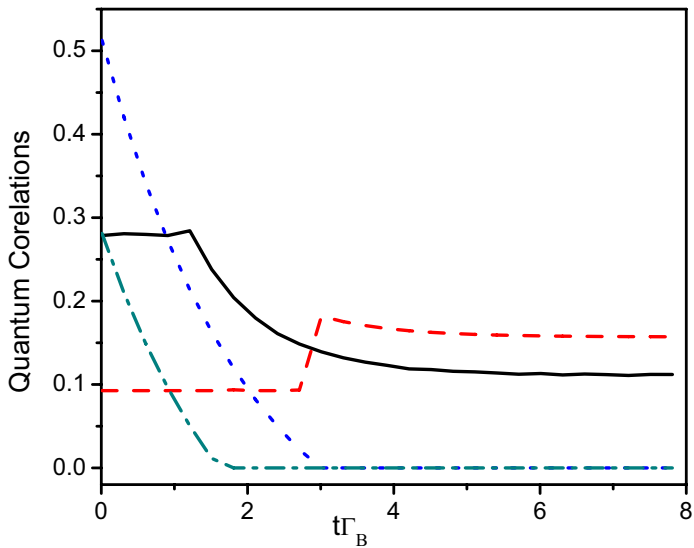}
\end{center}
\caption{(Color Online) Both the negativity (blue dotted, green dash-dotted)
and the quantum discord (black solid, red dashed) are plotted, respectively,
for two values of $p=(0.15,\hspace{2mm}0.23)$ against the dimensionless
parameter $t\Gamma _{B}$ for $t\Gamma _{A}=2$ when both the marginal systems
are coupled locally to dephasing noise.}
\label{Figure2}
\end{figure}

Figure \ref{Figure2} shows the dynamics of quantum correlations in
multilocal coupling for the same two values of $p$ against the dimensionless
parameter $t\Gamma_{B}$, which defines the qutrit's local environment.
Unlike the effect of qubit's environment on the dynamics of discord (solid
and dashed curves), we see that neither there is instantaneous decay nor the
decay is as fast as in figure \ref{Figure1}$b$. Most importantly, for $%
p=0.23 $ (dashed curve) the discord is static initially, then abruptly
increases, and remains almost static for the rest of the time. This behavior
shows the robustness of discord against the local coupling of the qutrit
marginal system with its environment. In fact, this is not only the case
with discord, a comparison with figure \ref{Figure1}$b$ shows that
entanglement (dashed and dash-dotted curves) is also comparatively robust in
this setup. We also confirm that our results reduce to the results of Ref. 
\cite{Karpat} for $\Gamma_{A}=0$.

\begin{figure}[h]
\begin{center}
\subfigure[]{
\includegraphics[scale=1.05]{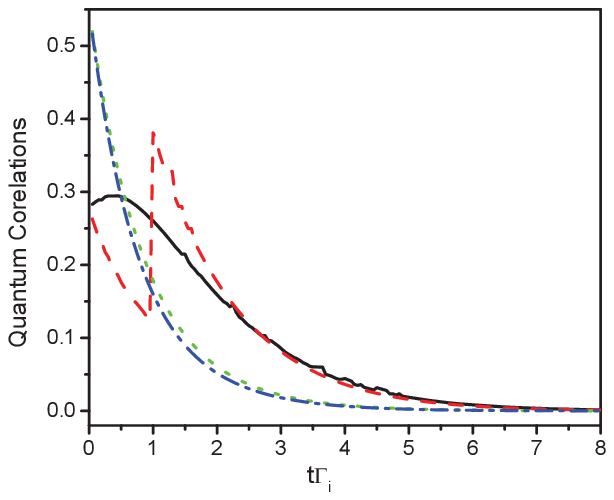}} 
\subfigure[]{
\includegraphics[scale=1.05]{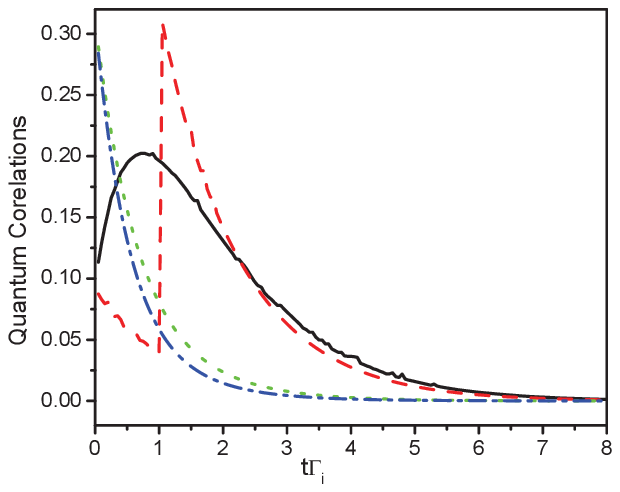}}
\end{center}
\caption{(Color Online) Both the negativity and the quantum discord are
plotted for $p=0.15$. (a) $p=0.23$ (b) against the dimensionless parameters $%
t\Gamma _{i}$ {$i=A,B$}. In both (a) and (b), The green (against $t\Gamma
_{A}$) and the blue (against $t\Gamma _{B}$) curves represent the behavior
of negativity and the black (against $t\Gamma _{A}$) and the red (against $%
t\Gamma _{B}$) curves represent the dynamics of quantum discord.}
\label{Figure3}
\end{figure}

\begin{figure}[h]
\begin{center}
\subfigure[]{
\includegraphics[scale=1.05]{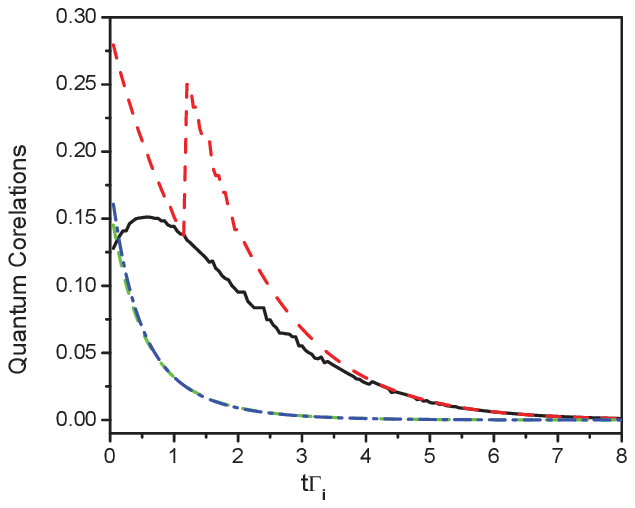}} 
\subfigure[]{
\includegraphics[scale=1.05]{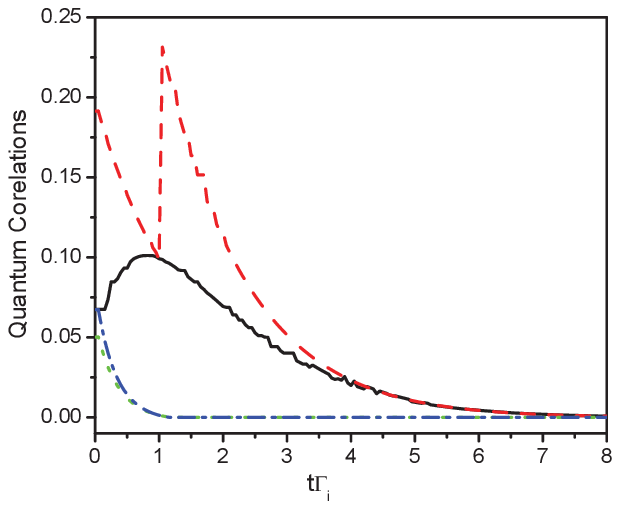}}
\end{center}
\caption{(Color Online) Both the negativity and the quantum discord are
plotted for $p=0.15$. (a) $p=0.23$ (b) against the dimensionless parameters $%
t\Gamma _{i}$ {$i=A,B$}. In both (a) and (b), The green (against $t\Gamma
_{A}$, $\Gamma _{B}=2$) and the blue (against $t\Gamma _{B}$, $\Gamma _{A}=2$%
) curves represent the behavior of negativity and the black (against $%
t\Gamma _{A}$, $\Gamma _{B}=2$) and the red (against $t\Gamma _{B}$, $\Gamma
_{A}=0.2$)curves represent the dynamics of quantum discord.}
\label{Figure4}
\end{figure}

Next, we numerically analyze the dynamics of entanglement and quantum
discord in the presence of amplitude damping noise. In figure \ref{Figure3}$%
a $, we show the dynamics of quantum correlations when only either the qubit
or the qutrit is locally coupled to its own environment. The green (dotted)
and the black (solid) curves represent the behaviors of entanglement and
quantum discord when only the qubit is locally coupled to its environment. 
The blue (dash-dot) and the red (dash) curves show the dynamics of
entanglement and quantum discord, respectively, when only the qutrit is
locally influenced by its environment. One can note that regardless of which
marginal system is coupled to its local environment, the loss of
entanglement is qualitatively identical. The same is true for quantum
discord in the limit of large values of the plotting parameters. However, in
the range of small values of $t\Gamma _{i}$, the behavior of quantum discord
strongly depends on which marginal system is coupled to its local
environment. The qubit's environment initially generates nonclassical
correlations in the system that grow smoothly and then monotonically decays
reaching zero in the asymptotic limit. On the other hand, the influence of
qutrit's environment causes nonanalytic changes in discord. Initially it
causes discord to monotonically decay reaching a minimum, abruptly increases
at a critical time reaching a maximum, again start decreasing monotonically
and then adopting behavior similar to the case of qubit coupling with its
environment as mentioned above. Since discord is a continuous function of
the plotting parameters, the abrupt increase in its dynamics at a critical
point may results due to the spontaneous generated coherence between the two
excited level of the qutrit system. The same argument applies to the
generation of dicord in figure \ref{Figure2} and figure \ref{Figure4} as
discussed next. Another notable aspect is the complete loss of quantum
discord in the asymptotic limit. The amplitude damping noise transforms the
initial quantum state into a classical-quantum state void of nonclassical
correlations when coupled with the system for long enough time. A more or
less similar behavior of quantum correlation is noted for $p=0.23$ which is
shown in figure \ref{Figure3}$b$. 
In figure \ref{Figure4}$(a,b)$, we plot the behaviors of quantum correlation
in the presence of multilocal environments for the same two values of $p$.
It can be seen that overall the qualitative behavior of quantum correlations
in the presence multilocal environments is unchanged, however, the peaks
height of quantum discord are considerably reduced. Secondly, for both
values of $p$ the quantum discord plotted against decay constant of qutrit's
environment is either large or equal to the one plotted against the decay
constant of qubit's environment.

\section{Summary}

In this paper we report the influence of noise on the dynamics of
entanglement and quantum discord in a one parameter qubit-qutrit systems. In
particular, we have considered the effects of phase damping and amplitude
damping noises in the Kruas operator formalism. Although the results are
discussed in detail in Section $\mathrm{IV}$, we summarize our findings in
this section. We have analytically demonstrated that both the one way and
two ways coupling of the environments with the system cause entanglement
sudden death. The quantum discord and hence the nonclassical correlations
survives when the system evolves under dephasing noise even in the
asymptomatic limit. We find that quantum discord not only show robustness in
coupling with qutrit's dephasing noise but can also be generated by the
qutrit's dephasing noise when each marginal system is locally coupled to its
own dephasing noise. On the other hand, it is found that in two ways
coupling discord is very fragile against the qubit's environment. Under such
coupling, the invariance and freezing behavior of quantum discord become
conditional. In the case of amplitude damping noise, the behavior of
entanglement is identical in both one way and two ways coupling of the
marginal systems. However, the dynamics of quantum discord strongly depends
on which marginal system is coupled locally to the environment. Amplitude
damping noise generates discord, however, in the case of qubit coupling with
environment, it is continuously generated whereas in the case of qutrit
coupling the generation results in a discontinuous manner.\newline

\end{document}